\documentclass[aps,twocolumn,floats,prd,nofootinbib,superscriptaddress,10pt]{revtex4-2}

\def\HI{{\rm H\,{\footnotesize I}~}}
\def\vcb{v_{cb}}
\def\Ts{T_{\rm S}}
\def\Tk{T_{\rm K}}
\def\Tcmb{T_{\rm CMB}}
\def\Tb{ T_{{\rm 21}}}
\def\vcb{v_{\rm cb}}
\def\lya{Lyman-$\alpha$~}
\def\mpci{~{\rm Mpc}^{-1}}
\def\apa{a_{\parallel}}
\def\ape{a_{\perp}}
\def\Msun{~{\rm M}_{\odot}}
\def\be{\begin{equation}}
\def\ee{\end{equation}}
\def\Trad{T_{\rm CMB}}
\def\kpa{k_{\parallel}}
\def\kpe{k_{\perp}}
\def\lx{\log_{10} \left( L_{\rm X}/{\rm SFR} \right)}

\usepackage{pifont}

\usepackage{amsmath}
\usepackage{amssymb}
\usepackage{graphicx}
\usepackage{multirow}
\usepackage{makecell}
\usepackage{hyperref}
\usepackage{adjustbox}
\hypersetup{
	colorlinks=true,
	linkcolor=red,
	filecolor=magenta,      
	urlcolor=blue,
}
\usepackage[normalem]{ulem}
\usepackage[usenames,dvipsnames]{xcolor} 
\usepackage{comment}
\usepackage{makecell}

\begin{document}


	\title{Measuring the cosmic expansion rate  using 21-cm velocity acoustic oscillations}

	\author{Debanjan Sarkar}
	\email{debanjan@post.bgu.ac.il}
	\affiliation{Physics Department, Ben-Gurion University of the Negev, Beersheba, Israel}

 	\author{Ely D. Kovetz}%
 	\email{kovetz@bgu.ac.il }
 	\affiliation{Physics Department, Ben-Gurion University of the Negev, Beersheba, Israel}

\begin{abstract}

The fluctuations in the dark matter-baryon relative velocity field are imprinted as 
acoustic oscillations in the 21-cm power spectrum during cosmic dawn (CD). 
These velocity acoustic oscillations (VAOs) keep the imprints of the comoving sound horizon scale.
In a previous work by Mu\~noz, it has been demonstrated that these VAOs can be treated as standard rulers
to measure the cosmic expansion rate at high redshifts by considering a variety of 
Lyman-Werner feedback strengths and foreground contamination scenarios. 
Here we extend that analysis by using a modified version of the public code \texttt{21cmFAST}.
We use this code to simulate the VAOs in 21-cm power spectrum
and forecast the potential to constrain $H(z)$ with the HERA radio telescope, 
taking into account the effects of Lyman-$\alpha$ heating, Lyman-Werner feedback and foregrounds,
the dependence on various astrophysical parameters, and the degeneracy with cosmological parameters.
We find that $H(z)$ can be measured with HERA at $\sim 0.3-6\%$ relative accuracy 
in the range $11 < z < 20$, under different astrophysical and foreground scenarios, 
with uncertainties in the Planck cosmological parameters  
setting a $\sim 0.08-0.2\%$ relative-error floor in the measurement.
This accuracy is on par with most low-redshift measurements and can be helpful in testing various cosmological scenarios motivated by the ongoing ``Hubble Tension".

\end{abstract}

\maketitle

\section{Introduction}
\label{sec:Introduction}


Measuring the cosmic expansion history is one of the key goals of modern cosmology. It started with 
the famous discovery of the expanding universe by Edwin Hubble about a century ago \cite{Hubble:1929ig}. 
Only about two decades ago, we came to know about the accelerating expansion of our universe
\cite{SupernovaSearchTeam:1998fmf,SupernovaCosmologyProject:1998vns}. 
Ever since this discovery,
a large number of theories have been proposed to explain this 
\cite{Weinberg:1988cp,Caldwell:2009ix,Joyce:2016vqv,Huterer:2017buf}. 
The theories can be
divided broadly into two classes: (i) theories that propose the existence of dark energy, which has a negative 
equation of state ($w$) and the cosmological constant ($\Lambda$) with $w=-1$ \cite{Carroll:2000fy}
being the simplest example of it,
(ii) theories that propose a modification in Einstein's theory of gravitation. Currently, we do not know 
which class of theories is correct, and we need to probe the cosmic expansion history
(which we also call as Hubble expansion and denote as $H(z)$) in order to pin down this.

We have measured the current expansion rate of the universe which is also known as the Hubble constant ($H_0$). 
Measurements of the Cosmic Microwave Background (CMB) using the Planck Satellite provide an 
extremely precise value of $H_0 =67.4 \pm 0.5$
km/s/Mpc \cite{Planck:2018vyg}. On the other hand, the local universe measurements like 
{\it SH0ES}, etc., which use observations of Cepheids in
the nearby galaxies, estimate the value of $H_0 = 73.0 \pm 1.0$ km/s/Mpc
\citep{Riess:2020fzl,Wong:2019kwg,DiValentino:2020zio}.
These two measurements are in $\sim 5-\sigma$ disagreement with each other, 
and this is coined as the Hubble “Tension”. 
For the time being, it is unclear whether this tension is caused by some new
physics beyond the standard cosmological model 
\cite{Knox:2019rjx,Jedamzik:2020zmd,Hill:2020osr,
Ivanov:2020ril,DAmico:2020kxu,Smith:2020rxx,Poulin:2018dzj,Smith:2019ihp,
Agrawal:2019lmo,Alexander:2019rsc,Lin:2019qug,Sakstein:2019fmf,Niedermann:2020dwg,
Kaloper:2019lpl,Berghaus:2019cls,Adi:2020qqf,Banerjee:2022ynv,Petronikolou:2021shp,DiValentino:2021izs}, or some
systematic effects \cite{DES:2022hav} in either or both of the measurements.
We, therefore, need other independent observations of the expansion rate in the local universe to 
draw a more robust conclusion on the cause of the $H_0$ tension, as well as at the high redshifts 
to test the cosmological theories of expansion.


The cosmological 21-cm signal will be observed over a large redshift range, as discussed in
Refs.~\cite{Madau:1996cs,Barkana:2000fd,Bharadwaj:2000av,Furlanetto:2006jb,Wyithe:2007gz,
Lewis:2007kz,Pritchard:2008da,Sarkar:2016lvb,Sarkar:2018gcb,Sarkar:2019nak,Sarkar:2019ojl}. 
Detection of the Baryon Acoustic Oscillation (BAO) features in the 21-cm intensity mapping power spectrum 
will provide the measurement of the expansion rate at $z \lesssim 6$
\cite{Bharadwaj:2008yn,Obuljen:2017jiy,Camera:2019iwy,Bernal:2019jdo}. 
BAOs that occur due to the interaction between the baryon and photon 
fluids before recombination, generate supersonic relative velocity between DM and baryons just after
recombination. This supersonic velocity prevents the formation of structures inside the 
mini halos (having masses in the range $10^5-10^7 \Msun$) \cite{Ciardi:2005gc,Kimm:2016kkj,Qin:2020xyh} 
that form early in hierarchical structure formation scenario
\cite{Bertschinger:1998tv,Strigari:2006jf,Somerville:2014ika}.
This delays the onset of the cosmic dawn (CD) epoch and directly affects
the evolution of the 21-cm signal \cite{Fialkov:2014rba, Barkana:2016nyr, 
Bovy:2012af, Tseliakhovich:2010bj, Stacy:2010gg, Fialkov:2011iw, Schmidt:2016coo, Munoz:2019rhi, Munoz:2019fkt}.
The fluctuations in the DM-baryon relative velocity field are imprinted as 
oscillations (just like BAOs), called velocity-induced acoustic oscillations (VAOs),
in the 21-cm power spectrum at large scales.
Ref.~\cite{Munoz:2019rhi} shows that these unique VAOs follow a simple analytic shape, 
which is set at the time of recombination.
Ref.~\cite{Munoz:2019fkt} first proposed to use the VAOs in the 21-cm power spectrum
as a standard ruler to measure the expansion 
rate at high redshifts.
Ref.~\cite{Munoz:2019fkt} showed that the 21-cm power spectrum measurements from the 
Hydrogen Epoch of Reionization Array (HERA) 
interferometer \cite{DeBoer:2016tnn}
should be able to measure the Hubble expansion rate $H(z)$ at $z = 15 - 20$ to percent-level precision, 
depending on the strength of Lyman-Werner (LW) feedback process \cite{Visbal:2014fta,Ricotti:2000at,Haiman:1996rc}
and foreground contamination.

The analysis in Ref.~\cite{Munoz:2019fkt} is done 
at two fixed redshifts, and
for a single set of 
cosmological and astrophysical parameters.
Note that the onset of the CD, as well as the amplitude of the 21-cm signal depends on the 
cosmological and astrophysical parameters. For fixed sensitivity of telescopes
and foreground contamination scenario, 
the amplitude of the 21-cm signal (or of the VAO peaks) mainly decides 
the signal-to-noise (SNR) with which VAOs can be detected. 
The best SNR occurs at redshifts which is determined by the combination of the model parameters 
and the telescope sensitivity. Note that, the binning of the power 
spectrum is also important as we need certain number of 
bins to clearly make out the VAO features which is important for the $H(z)$ measurement, and
the VAO features will get smoothed out if less number of bins are used. 
Therefore, considering observations with a fixed telescope, 
we need to explore the effects of the different choice of the model parameters on the measurement
of $H(z)$.  
The measurement of $H(z)$ is further subject to the additional heating effects like the \lya heating 
which raise the IGM temperature
if the poorly-constrained X-ray heating is not extremely efficient.
The \lya heating is due to the resonant scattering between \lya photons and the IGM 
atoms~\cite{Chuzhoy:2006au, Chen:2003gc, Oklopcic:2013nda, Ciardi:2009zd, Mittal:2020kjs}.
\footnote{Note that, there is another possible heating mechanism, called  CMB heating, 
which results from the energy transfer from the radio background 
(which is dominated by the CMB) into the IGM, mediated by the \lya photons \cite{Venumadhav:2018uwn}.
Some recent works debate the significance of this effect~\cite{Meiksin:2021cuh}, and
we do not include CMB heating in our current analysis. We note that, our conclusions are not very sensitive 
to this, as the \lya heating alone accounts for most of the heating effect.}
In Ref.~\cite{Sarkar:2022dvl}, we showed that \lya heating suppresses the 21-cm power spectrum
amplitude and this will likely affect the detectability of the VAOs.

Although the cosmological parameters are well measured by the Planck 
mission \cite{Planck:2018vyg}, there are still small uncertainties in the measured parameters.
We need to estimate the errors propagated into the $H(z)$ measurement
due to the uncertainties in the cosmological parameters.


In this paper, we simulate the 21-cm signal and 
model the VAOs in the 21-cm power spectrum using analytical prescriptions. 
Considering observations with the HERA radio telescope, 
we quantify the relative error with which the expansion rate $H(z)$
can be measured under the different heating, LW feedback and foreground contamination scenarios. 
For each simulation scenario, we contained our analysis only to EoH, 
which occurs at $z<z_{\rm min}$ (where $z_{\rm min}$ 
refers to the redshift where the global 21-cm signal 
$\langle T_{21} \rangle = \langle T_{21} \rangle_{\rm min}$ has its minimum
value), and measure $H(z)$ at $z=z_{\rm half}$ where the global signal is
$\langle T_{21} \rangle = (1/2) \langle T_{21} \rangle_{\rm min}$ and the signal-to-noise for VAO 
measurement is expected to be the highest. 
We also quantify the errors in $H(z)$ measurement introduced by the 
uncertainties in cosmological parameters from Planck measurements.
The paper is organized as follows.

In Section~\ref{sec:formalism}, we discuss the 21-cm signal from different epochs, briefly review the
modelling of the DM-baryon relative velocity effect in the 21-cm power spectrum, and 
outline the procedure to measure the expansion rate $H(z)$ from the VAOs. In Section~\ref{sec:simulation},
we discuss our simulations and model parameters. In section~\ref{sec:sensitivity}, we present the
sensitivity calculation for HERA radio telescope. We present our main findings in Section~\ref{sec:results},
and conclude in Section~\ref{sec:summary}. Note that, throughout our analysis, we have assumed flat $\Lambda$-CDM
cosmology and use the fiducial cosmological parameters as given in Table~\ref{tab:parameters-fid}.

\section{Formalism}
\label{sec:formalism}

\subsection{21-cm Signal from Different Epochs}
\label{sec:epochs}

The 21-cm brightness temperature is given by \cite{Bharadwaj:2000av,Furlanetto:2006jb}
	\begin{equation}
		\Tb=\frac{\Ts - \Trad}{1+z} \left( 1 - e^{-\tau_{21}}\right) \,,
		\label{eq:T21}
	\end{equation}
	where $\Ts$ is the spin temperature, $\Trad$ is the temperature of the background radiation which is usually
	assumed to be CMB with $\Trad = T_{\rm CMB}(z) = 2.7255 (1+z) \,{\rm K}$, and $\tau_{21}$
	is the 21-cm optical depth which can be calculated as
	\begin{equation}
		\tau_{21}=\frac{3hA_{10}c\lambda_{21}^{2}n_{{\HI}}}{32\pi k_{{\rm B}}T_{{\rm S}}(1+z)(dv_{r}/dr)}\,.  
		\label{eq:tau21}
	\end{equation}
	Here, $h$ is the $Planck$ constant, $A_{10}$ is the Einstein $A$-coefficient for the 21-cm emission, $c$ is the speed of light,
	$\lambda_{21}$ is the wavelength of the 21-cm radiation, $n_{\HI}$ is the neutral hydrogen number density, $k_B$ is 
	Boltzmann constant, $dv_{r}/dr$ is the gradient of the comoving velocity along the line of sight which is taken to be 
	$H(z)/(1+z)$ where $H(z)$ is the Hubble rate. 
	
	The spin temperature can be calculated as \cite{Venumadhav:2018uwn,Sarkar:2022dvl}
	\begin{equation}
		\Ts=\frac{x_{{\rm rad}} + x_{\alpha} + x_c}{x_{{\rm rad}}T_{{\rm rad}}^{-1} + x_c T_{{\rm K}}^{-1} + 
			x_{\alpha} T_{\rm c,eff}^{-1} }\,,
		\label{eq:Tspin}
	\end{equation}
	where,
	\begin{equation}
		x_{{\rm rad}}=\frac{1-e^{-\tau_{21}}}{\tau_{21}} \,,
		\label{eq:xrad}
	\end{equation}
	$x_{\alpha}$ and $x_c$ are Wouthuysen-Field coupling \cite{Wouthuysen:1952,Field:1958,Hirata:2005mz}
	and Collisional \citep{Furlanetto:2006su} 
	coupling coefficients respectively, and $T_{\rm c,eff}$
	is the effective colour temperature \cite{Furlanetto:2006jb} for the \lya radiation.

The standard evolution history of the 21-cm line is
as follows. During the dark ages \cite{Shapiro:2005cx}, collision coupled the spin temperature with the gas, 
producing 21-cm absorption for $z \gtrsim 30$. Collision, however, became inefficient as the universe 
expanded and the gas cooled and diluted further. This resulted in nearly no absorption until the 
advent of CD epoch ($z\sim30$ for all the curves in Fig.~\ref{fig:global_sig})
\cite{Mellema:2012ht,Ocvirk:2015xzu,Park:2018ljd}. 
In this epoch the first luminous sources were formed. 
These sources emitted ample ultraviolet (UV) photons, which coupled $\Ts$ to $\Tk$ through a process called 
the Wouthuysen-Field (WF) effect \cite{Wouthuysen:1952,Field:1958,Hirata:2005mz}, 
by which \lya photons resonantly scatter between hydrogen atoms, 
imprinting $\Tk$ onto the hyperfine populations of hydrogen. This again produces 21-cm absorption and 
we call this period the Lyman-$\alpha$ coupling era (LCE) (which ranges from $z \lesssim 30$
to $z\sim 20$ for the solid curves in Fig.~\ref{fig:global_sig}). 
Later on, X-ray \cite{Ciardi:2009zd,Xu:2014aja,Ewall-Wice:2015uul,Sazonov:2016vac,Sazonov:2016vac} 
and other photons, like \lya \cite{Chen:2003gc,Chuzhoy:2006au}
, re-heat the IGM to temperatures 
$\Tk \gtrsim \Tcmb$, and switched the 21-cm signal to emission
($z<12$ for the solid curves in Fig.~\ref{fig:global_sig}). 
We call this period the epoch of heating (EoH)
(For the solid curves in Fig.~\ref{fig:global_sig}, it ranges from $z\lesssim 20 $ to the beginning of reionization.
The reionization for the curves considered began at $z\sim10$ when the 21-cm global signal start to decline due to the
lack of neutral hydrogen fraction in the IGM). Eventually at $z<10$, i.e., at epoch of reionization (EoR) \cite{Benson:2005cc}, 
high amount of UV photons ionized most of the neutral hydrogen that almost no 21-cm signal is left.
Note that the minimum of the global signal, which is $\langle T_{21} \rangle_{\rm min} \sim -165$ mK
at $z\sim 19.5$ for the blue curve in Fig.~\ref{fig:global_sig}, marks the end of LCE and beginning of EoH.
The redshift, where this happens, we denote it by $z_{\rm min}$.

The effect of relative velocities on the 21-cm brightness temperature during the LCE and EoH are different
(see Ref.~\cite{Munoz:2019rhi} for a comprehensive discussion on this). 
In LCE, the regions with large velocity produce shallower 21-cm absorption as the velocities impede star formation
and thereby reduce \lya photon production. Regions with small velocity produce deeper 21-cm absorption
as the \lya photon production is not hampered by much. This effect is most prominent at the redshift 
where $\langle T_{21} \rangle = (1/2) \langle T_{21} \rangle_{\rm min}$ ($z\sim 25$ for the solid curves in Fig.~\ref{fig:global_sig}) and we denote this redshift as $z^{\rm LCE}_{\rm half}$. This effect fades away 
gradually as we go towards the lower redshift where the \lya coupling saturates. In EoH, the opposite happens.
Patches with large velocities form fewer stars, thereby producing less heating,
and therefore causing deeper 21-cm absorption. Patches with small velocities producing more heating,
and therefore causing shallower 21-cm absorption. This effect also is most prominent at the redshift 
where $\langle T_{21} \rangle = (1/2) \langle T_{21} \rangle_{\rm min}$ ($z\sim 15$ for the blue curve in Fig.~\ref{fig:global_sig}) and we denote this redshift as $z^{\rm EoH}_{\rm half}$. This effect becomes less
important as the IGM heating causes $\Ts \gg \Tcmb$ and the fluctuation due to $\Ts$
is negligible in Eq.~\eqref{eq:T21}. Due to the opposite nature at LCE and EoH,
we have practically no effect of velocities at $z_{\rm min}$. 
Ref.~\cite{Munoz:2019rhi} showed that the VAO amplitude in the 21-cm power spectrum is most prominent 
at $z^{\rm LCE}_{\rm half}$ and $z^{\rm EoH}_{\rm half}$, and 
the VAO amplitude is decreased as we move away from these redshifts. Based on the above discussion, 
we have no clear VAOs at $z_{\rm min}$, as well as at redshifts where $\Ts \gg \Tcmb$
\cite{Fialkov:2013uwm}. The VAOs also
have higher amplitude during the EoH.   
Ref.~\cite{Munoz:2019rhi} also showed that the signal-to-noise for the VAO detection with the current generation of 21-cm
experiments, like HERA, is higher during the EoH and expected to be maximum close to $z^{\rm EoH}_{\rm half}$.
For this reason, from now onwards, we focus on EoH and we denote 
$z_{\rm half} \equiv z^{\rm EoH}_{\rm half}$ for brevity.

\subsection{Model the dark matter-baryon relative velocity effects in the 21-cm Power spectrum}
In this section, we discuss the modulation of the 21-cm power-spectrum due to the DM-baryon
relative velocity.
The 21-cm power-spectrum is defined as,
\begin{equation}
\Delta^2_{21}(k)=\frac{k^3P_{21}(k)}{2 \pi^2} \,[{\rm mK}^2],
\label{eq:power-spec}
\end{equation}
where $P_{21}(k)=\langle \Tilde{T}_{21}(k) \Tilde{T}_{21}^{\ast}(k) \rangle$ and $\Tilde{T}_{21}(k)$ is the 
Fourier transform of $T_{21}-\langle T_{21}\rangle$.
The modulation of the 21-cm power spectrum due to the
dark-matter–baryon relative velocities can be apprehended
from the statistics of the collapsed
baryonic density. The effect of bulk relative velocities is very similar to that of the baryonic pressure, 
which suppresses the accumulation of baryons in the haloes. As the gas accreted into the DM halo, 
the relative velocity increases the effective sound
speed. This further increases the critical mass scale of a halo that can retain the baryons and
decreases the baryon collapsed fraction
\cite{Dalal:2010yt,Naoz:2011if,Greif:2011iv,McQuinn:2012rt,Stacy:2010gg,Fialkov:2011iw,Yoo:2011tq,
Pritchard:2011xb,Barkana:2016nyr,Tseliakhovich:2010yw}.
The effect of the relative velocities on the amplitude of
the 21-cm brightness temperature power spectrum can
be parameterised as \cite{Munoz:2019rhi}
\be
\Delta_{21,\rm vel}^2(k,z)=A_{\rm vel}(z)\Delta_{v^2}^2(k,z)|W(k,z)|^2\,,
\label{eq:pk-VAO}
\ee
where $A_{\rm vel}$ is some redshift-dependent amplitude of fluctuations.
The window function $W(k,z)$ depends on the different contributors to the 21-cm power spectrum such as 
the coupling to the Lyman-$\alpha$ photons and $X$-ray heating. 
Here, $\Delta_{v^2}^2(k)$ as the power spectrum of the quantity 
\be
\delta_{v^2}=\sqrt{\frac{3}{2}}\left(\frac{v_{\rm cb}^2}{v_{\rm rms}^2}-1\right)\,,
\label{eq:vsq}
\ee
which accurately captures the shape of the effect of relative velocities on the observables 
for the scales of interest and the `streaming' bulk relative velocity ($\vcb$) 
can be approximated with a root-mean-squared value $v_{\rm rms}\simeq30 \,{\rm km\,s}^{-1}$ at recombination~\citep{Tseliakhovich:2010bj}. Note that, $A_{\rm vel}$ is a model-dependent 
amplitude that is not directly observable. The VAOs are statistically independent 
from the density fluctuations at first order.
Therefore, the amplitude of the 21-cm power spectrum can be written as \cite{Munoz:2019rhi}
\be\label{eq:vao_definition}
\Delta_{21}^2(k,z)=\Delta_{21,\rm vel}^2(k,z)+\Delta^2_{21,\rm nw}(k,z)\,,
\ee
where $\Delta^2_{21\!,\rm nw}(k,z)$ is the component of the 21-cm power spectrum \textit{without} VAOs. 
Following Ref.~\cite{Munoz:2019rhi},  we parameterise the $\Delta^2_{21\!,\rm nw}(k,z)$ as a smooth 
polynomial,
\be\label{eq:smooth_pw}
\ln[\Delta_{21,\rm nw}^2(k,z)]=\sum_{i=0}^{n}c_i(z)[\ln k]^i\,,
\ee
where $c_i(z)$ are coefficients we fit for using simulations as discussed in Ref.~\cite{Munoz:2019rhi}. 
We model the velocity power spectrum as in Ref.~\citep{Munoz:2019fkt} using the form we defined 
in Eq.~\eqref{eq:pk-VAO}. 
We calculate the window function and the amplitude $A_{\rm vel}(z)$ for a given model using  
\texttt{21cmvFAST}~\cite{Munoz:2019rhi} (which we shall discuss later), 
and calculate $\Delta_{v^2}^2$ for a given cosmology. 
We calculate the transfer function of the relative velocities at the end of recombination using 
\texttt{CLASS} Boltzmann code \cite{Lesgourgues:2011re,Blas:2011rf,Lesgourgues:2011rg,Lesgourgues:2011rh}, 
which is then used as an initial condition for our simulations.

\subsection{Measurements of $H(z)$ from acoustic peaks}
\label{sec:Hz_measurement}

Since  VAOs are sourced by BAOs, they keep the imprints of the comoving sound horizon scale
$r_d \approx 150$ Mpc at the baryon drag era ($z_d \approx 1060$). The separation of the VAO peaks 
in the Fourier-space, $\Delta k = 2 \pi/r_d$, gives the estimate of $r_d$ and the VAO features in 
the CD 21-cm power spectrum preserves this well known distance scale \cite{Munoz:2019fkt}. 
This $r_d$ can be used as a 
standard ruler, and together with the Alcock-Paczy\'{n}sky (AP) 
\cite{Alcock:1979mp,Barkana:2005nr,Li:2019ilp,Bernal:2019gfq,Melia:2020hur}
test on the power spectrum data, we can 
recover the expansion rate $H(z)$. According to AP effect, the parallel $k_{\parallel}$
and perpendicular ${\bf k}_{\perp}$ wave vectors (defined with respect to the line-of-sight 
direction) get shifted to values $k_{\parallel}/\apa$ and ${\bf k}_{\perp}/\ape$ 
when assuming wrong fiducial cosmology \cite{Bernal:2019gfq}. 
Here $\apa = (H^{\rm fid}(z) r^{\rm fid}_d)/(H(z)r_d)$
and $\ape = (D_A(z) r^{\rm fid}_d)/(D^{\rm fid}_A(z) r_d)$ are two AP parameters, where $H(z)$ is the
Hubble expansion rate, $D_A(z)$ is the angular diameter distance, and superscript ``fid" denotes their 
fiducial values. Therefore, measuring the shift in the VAO peaks, we can constrain $\apa$ and 
$\ape$, and thereby measure $H(z)$ and $D_A(z)$ \cite{Munoz:2019fkt}. 

\section{Simulation}
\label{sec:simulation}
	
We use a modified version of the publicly available
 \texttt{21cmvFAST}\footnote{\href{https://github.com/JulianBMunoz/21-cmvFAST}
{github.com/JulianBMunoz/21-cmvFAST}}
\cite{Munoz:2019rhi}
semi-numerical code to generate the observable 21-cm signal. 
\texttt{21cmvFAST} mainly includes
the effects of DM-baryon relative velocity and LW radiation 
feedback into the 21-cm calculations, using pre-calculated input tables of quantities that depend on 
these effects, given for
a single set of cosmological parameters (matching Planck cosmology). 
Note that, \texttt{21cmvFAST} itself is a modification of the public code \texttt{21cmFAST} 
\footnote{\href{https://github.com/21cmfast/21cmFAST.git}
{github.com/21cmfast/21cmFAST.git}} \cite{Mesinger:2010ne}, widely used to simulate the 21-cm signal.
We modify \texttt{21cmvFAST} in order to interface it with {\tt CLASS}
\cite{Lesgourgues:2011re,Blas:2011rf,Lesgourgues:2011rg,Lesgourgues:2011rh},
which enables it to calculate all required quantities on the fly for any cosmological scenario 
and any set of input cosmological parameters. We have also introduced the CMB and \lya heating effects
in the code. The details about these implementations can be found in Ref.~\cite{Sarkar:2022dvl}.
	
The \texttt{21cmvFAST}~\cite{Munoz:2019rhi} code requires a number astrophysical and cosmological parameters
as input. 
The astrophysical
parameters are: $\zeta$ (describes the efficiency of ionizing photon production),
$\lambda_{\rm MFP}$ (mean free path of the ionizing photon),
$V^{(0)}_{\rm cool}$ (minimum halo mass for  molecular cooling in the absence of relative velocity),
$V^{\HI}_{\rm cool}$ (minimum halo mass for atomic cooling),
$\log_{10} \left( L_{\rm X}/{\rm SFR} \right)$ (log of X-ray luminosity, normalized by the star formation rate SFR, in units of $\mathrm{erg}\,\,\mathrm{s}^{-1}\,\,M_\odot^{-1}\,\,\mathrm{yr}$),
$\alpha_X$ (X-ray spectral index),
$f^0_{\star}$ (fraction of baryons in stars),
$E_{\rm min}$ (threshold energy, below which we assume all X-rays are self-absorbed near the sources).
	
We assume a flat universe with the following cosmological parameters: 
$h$ (Hubble parameter),
$\sigma_{8,0}$ (standard deviation of the current matter fluctuation smoothed at scale $8\,h^{-1}$Mpc)
$\Omega_{\rm m0}$ (total matter density at present),
$\Omega_{\rm b0}$ (total baryon density at present),
$n_s$ (spectral index of the primordial power spectrum),
$T_\mathrm{CMB}$ (current CMB temperature). The fiducial values of these parameters are given in 
Table~\ref{tab:parameters-fid}.
	
We run our modified version of \texttt{21cmvFAST} with box sizes $600$ Mpc and $1$ Mpc resolution to compute the 21-cm global signal and fluctuations. We verified that the choice of a $600$ Mpc box retains
sufficient $\vcb$ power at large scales and the power spectra show good convergence with a $900$ Mpc box
results. For a better visualization of the VAO shape, we increase the number of bins while 
calculating the 21-cm power spectrum. This, however, decreases the number of ${\bf k}$-modes
and increases the Poisson noise in each bin.

\begin{table}[]
	\begin{tabular}{|c|c|}
		\hline
		Parameters & Fiducial Values \\  \Xhline{4\arrayrulewidth}
		$\zeta$ & $20$    \\ \hline
		$\lambda_{\rm MFP}$ & $15$ Mpc  \\ \hline
		$V^{(0)}_{\rm cool}\, [{\rm km/s}]$ & 4  \\ \hline
		$V^{\HI}_{\rm cool}\, [{\rm km/s}]$ & 17  \\ \hline
		$\log_{10} \left( L_{\rm X}/{\rm SFR} \right)$ & $39$  \\ \hline
		$\alpha_X$ & $1.2$  \\ \hline
		$f^0_{\star}$ & $0.05$  \\ \hline
		$E_{\rm min}$ & $0.2$ keV \\  \Xhline{4\arrayrulewidth}
		$\sigma_{8,0}$ & 0.8102  \\ \hline
		$h$ & 0.6766   \\ \hline
		$\Omega_{\rm m0}$ & 0.3111   \\ \hline
		$\Omega_{\rm b0}$ & 0.0489  \\ \hline
		$n_s$ & 0.9665  \\ \hline
		$T_{\rm CMB}$ & 2.7255  \\  \Xhline{4\arrayrulewidth}
					
	\end{tabular}
	\caption{Our main simulation parameters and their fiducial values.}
	\label{tab:parameters-fid}
\end{table}
	
We  have considered three LW radiation feedback strengths in our simulations
(i) no feedback, (ii) low feedback and (iii) regular feedback, as defined in Ref.~\cite{Munoz:2019rhi}.
Further, we only consider  \lya heating in our analysis due to the uncertainties in the CMB heating
efficiency as discussed in Section~\ref{sec:Introduction}. For weak X-ray efficiency, 
Ref.~\cite{Sarkar:2022dvl} showed that \lya heating dominates most of the EoH, 
and our results largely do not depend on the CMB heating. 
In our fiducial simulations, we  consider $\lx=39$. 
Recently, Ref.~\cite{HERA:2021noe} showed that the HERA Phase-I data \cite{HERA:2021bsv}
suggests high X-ray efficiency ($\lx>40$) for high redshift galaxies. However, the
analysis in Ref.~\cite{HERA:2021noe} does not incorporate \lya heating, which will presumably
bring down the preferred $\lx$ value. Based on this, we set $\lx=39$
to compensate for the extra heating caused by \lya photons.
 Note that, for high X-ray efficiency, the \lya heating is subdominant \cite{Sarkar:2022dvl}.

As defined earlier, $\Delta_{21,\rm nw}^2(k,z)$ is the power spectrum of the 21-cm signal without the 
VAOs. VAOs result from the fluctuations in the $\vcb$ field. To suppress the  fluctuations in the $\vcb$ field,
we replace the $\vcb$ values with $\vcb=\langle \vcb \rangle = 0.92 \, v_{\rm rms}$ in the simulations, as suggested
in Ref.~\cite{Munoz:2019rhi}. This choice only suppresses the VAOs in 
the 21-cm power spectrum, without altering the 
global signal.

\section{Sensitivity Calculation for HERA}
\label{sec:sensitivity}
	
	The measurement of $H(z)$ depends on the detectability of the  of the VAO peaks 
	and their shift. The detectability, however, depends on the sensitivity of the 
	radio telescope and the foreground contamination. In the 21-cm observations, the foreground is
	several order of magnitude brighter than the signal and
	expected to contaminate a significant amount of Fourier space
	\cite{Bowman:2008mk,Dillon:2012wx,Hazelton:2013xu,Liu:2011hh}. 
	In the $\kpe-\kpa$ space, where $\kpa$ and $\kpe$ are the components of the wave vector 
	respectively parallel and perpendicular
	to the line-of-sight direction, the contaminated part of the 
	Fourier space looks like a ``wedge" \cite{Datta:2010pk,Pober:2013jna}. 
	Following Refs.~\cite{Pober:2012zz,Pober:2013jna}  
	the extent of the foreground wedge can be parametrized by assuming that all 
	wave numbers with $\kpa$ below
	\be
	\kpa^{\rm min} = a + b(z) \kpe
	\ee
	are contaminated, where $b(z)$ accounts for the chromaticity of the antennae, 
	and $a$ is a constant superhorizon buffer.
	Given the uncertainty in foreground contamination, we consider three cases \cite{Munoz:2019fkt}.
	In the ``optimistic" case, where we assume minimum amount of foreground contamination,
	we set $a=0$ and $b(z)$ is determined by the primary beam. In the ``moderate" 
	and ``pessimistic" foreground contamination scenarios, we assume $b(z)$ is determined 
	by the horizon limit, and consider $a = (0.05,0.1)\,h\,\mpci$ respectively. Further more, 
	the baselines are coherently added in case of optimistic and moderate foregrounds, while
	only the instantaneously redundant baselines are combined coherently in the pessimistic case.

	For the forecast, we consider the HERA 21-cm intensity mapping experiment~\cite{DeBoer:2016tnn}. 
	HERA is located in the Karoo Desert of South Africa and is designed to measure the 21-cm fluctuations 
	from CD ($50$ MHz or $z\sim27$) to the reionization era ($225$ MHz or $z\sim5$). 
	The final stage of HERA is expected to have $350$
	antenna dishes, each with a diameter of $14$ m. Out of the $350$ dishes, $320$ will be 
	placed in a close-packed hexagonal configuration and the remaining $30$ will be placed at longer 
	baselines. We calculate the sensitivity of HERA using the publicly available
	package \texttt{21cmSense}
	\footnote{\href{https://github.com/steven-murray/21cmSense}{github.com/steven-murray/21cmSense}}
	~\cite{Pober:2012zz,Pober:2013jna}. 
	This code accounts for the $u-v$ sensitivities of each antenna
	in the array, and calculates the possible errors in the 21-cm power spectrum measurement,
	including cosmic variance. For details, the reader is referred to Ref.~\cite{Sarkar:2022dvl}. 
	We consider a total of $540$ days of observation per redshift with 6 hours of observation per day. 
	Following Ref.~\cite{Munoz:2019fkt}, we bin the $k$-modes logarithmically in order to resolve the VAO peaks
	more clearly.

\section{Results}
\label{sec:results}

\subsection{Effect of Lyman-$\alpha$ Heating}
\label{sec:effect_lya_heat}

\begin{figure}
    \centering
    \includegraphics[width=1\columnwidth]{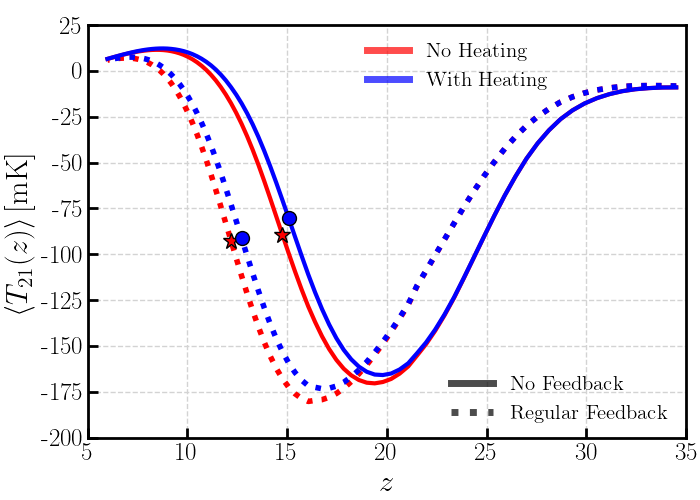}
    \caption{This shows the global 21-cm signal $\langle T_{21} \rangle$ for our fiducial parameters
    and for two LW feedback scenarios. The solid curves show signal for no feedback, while 
    the dashed curves show signal for regular feedback. The blue and red colours respectively indicate 
    simulations with and without \lya heating. Filled stars and filled circles represent the redshift
    values (which we denote by $z_{\rm half}$, and the values are given in Table~\ref{tab:table1})
    where $\langle T_{21} \rangle = (1/2) \langle T_{21} \rangle_{\rm min}$.
    Here, $\langle T_{21} \rangle_{\rm min}$ is the minimum value of the global signal which occurs at a
    redshift which we call $z_{\rm min}$. The LW radiation feedback disturbs the star formation in the 
    haloes, which delays the cosmic dawn epoch and all the epochs that follow. This can be seen clearly.
    Comparing solid and dashed curves, we see that all the different epochs like LCE, EoH (discussed in Section~\ref{sec:epochs}) are shifted towards the low redshifts by $\Delta z \sim 3$
    when we introduce LW feedback. The \lya heating
    heats up the IGM and, as a result, the EoH starts early for the models with \lya heating. The 
    blue curves show higher $\langle T_{21} \rangle_{\rm min}$ than the red curves. Also the $z_{\rm min}$ and
    $z_{\rm half}$ values are slightly higher for the models with \lya heating. }
    \label{fig:global_sig}
\end{figure}

 In this section, we discuss the effects of \lya heating on the 21-cm global signal and power spectrum.
 Figure~\ref{fig:global_sig} shows the 21-cm global signal for our fiducial model parameters 
 (see Table~\ref{tab:parameters-fid})
 and for two different LW feedback scenarios: no feedback and regular feedback. We find that the 
 cosmic dawn and the subsequent epochs are delayed (by  $\Delta z \sim 3$)
 for the regular feedback cases, compared to
 the no feedback cases. As a result, $\langle T_{21} \rangle_{\rm min}$ occurs at lower $z_{\rm min}$ 
 (making $z_{\rm half}$ lower as well)
 for the models with regular feedback. Also, $\langle T_{21} \rangle_{\rm min}$ values are more negative for the 
 models with regular feedback. Now considering a particular feedback case, 
 we see that $\langle T_{21} \rangle_{\rm min}$
 value is raised by a little and occurs at a slightly higher $z_{\rm min}$ when we introduce the \lya
 heating. In the \lya heating scenario, the \lya photons heat the IGM and raise the kinetic temperature 
 $\Tk$. As a result, contrast between $\Tcmb$ and $\Ts$ is decreased and this increases 
 $\langle T_{21} \rangle_{\rm min}$. The $z_{\rm half}$ values are also slightly higher for the models with \lya
 heating, compared to no heating. The $z_{\rm half}$ values for different LW feedback strengths and \lya heating
 scenarios are given in Table~\ref{tab:table1}.
 Note that, the \lya heating is more efficient for the models with 
 low X-ray efficiency. In our fiducial models with $\lx=39$, we have sufficient contribution from the 
 \lya heating. We find that for models with $\lx > 40$, the \lya heating contribution is negligible. 
 Overall, \lya heating tries to decrease the contrast between $\Tcmb$ and $\Ts$ and this has a
 significant effect on the 21-cm power spectrum, which we shall see next.

 \begin{figure}
    \centering
    \includegraphics[width=1\columnwidth]{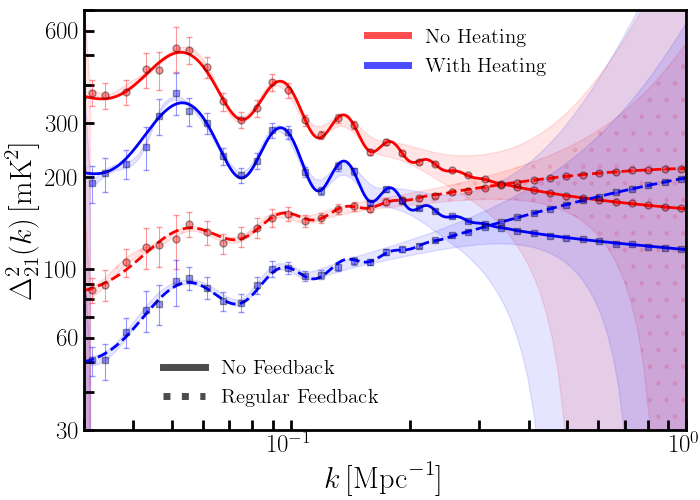}
    \caption{This shows the 21-cm power spectrum $\Delta^2_{21}(k)$ as function of 
    wave vector $k$ at redshifts $z_{\rm half}$ where the global signal is 
    $\langle T_{21} \rangle = (1/2) \langle T_{21} \rangle_{\rm min}$ (shown using
    filled stars and circles in Fig.~\ref{fig:global_sig}). Here we show the 
    same simulations as in Fig.~\ref{fig:global_sig} and all the different lines 
    and colours carry the same meaning. Here the simulated data are shown with points and the error
    bars show the $1-\sigma$ Poisson error at each bin. 
    The shaded regions show errors obtained for HERA using \texttt{21cmSense} under the optimistic
    foreground assumption.
    The solid and dashed lines represent the fits
    of the model given in Eq.~\eqref{eq:vao_definition}. We see that feedback reduces the amplitude of the
    power spectrum and the associated VAO peaks. LW feedback disturbs the star formation in the smaller
    haloes and decreases the fluctuations in the 21-cm field introduced by the relative velocity.  
    Although the $z_{\rm half}$ values are smaller for the 
    regular feedback cases, compared to no feedback cases, the suppression is significant. 
    Similarly, \lya heating also decreases the amplitude of the 21-cm power spectrum and VAOs. 
    The \lya heating increases the IGM temperature, and thereby decreasing the contrast between 
    $\Tcmb$ and $\Ts$. }
    \label{fig:pk_vao}
\end{figure}

 Figure~\ref{fig:pk_vao} shows the 21-cm power spectrum $\Delta^2_{21}(k)$ at $z=z_{\rm half}$ 
 (see Table~\ref{tab:table1})
 for the same simulations shown in Figure~\ref{fig:global_sig}. We see that both feedback and
 \lya heating decrease the amplitude of the 21-cm power spectrum and VAO peaks. The 
 difference $\Delta z_{\rm half} \sim 3$ between regular and no feedback cases is large, 
 and we cannot directly
 compare the feedback effects as they occur at very different redshifts. However, previous studies 
 (like Ref.~\cite{Sarkar:2022dvl})
 showed that for a fixed $z$, LW feedback suppresses the amplitude of the power spectrum significantly.
 On the other hand, the difference $\Delta z_{\rm half}\sim 0.3-0.5$ between with and without \lya
 heating scenarios is small, and we can directly compare the heating effects here. We see that \lya heating suppresses the
 VAO peaks by a factor of $\sim 2$. Not only that, it pushes $z_{\rm half}$ at a higher value. Both of these effects,
 influence the effective signal-to-noise \citep{Munoz:2019rhi} with which VAO peaks can be detected, and
 affect the measurement of $H(z)$, which we shall see later.

\subsection{Relative error on $H(z)r_d$ measurement}
\label{sec:measurement}

In this section, we discuss the relative error on the measurement of $H(z)$ based on the 
method outlined in Section~\ref{sec:Hz_measurement}.
We generate the mock data, which is the 21-cm power spectrum $\Delta^2_{\rm 21,data}$, using 
the \texttt{21cmvFAST} simulation (discussed in Section~\ref{sec:simulation}). 
For the first part of our analysis, we generate data using our
fiducial model parameters (Table~\ref{tab:parameters-fid}) 
for three different LW feedback strengths and \lya heating
(as given in Table~\ref{tab:table1}). We define our likelihood $\mathcal{L}$ at each redshift as
\be
-\log \mathcal{L} = \frac{1}{2} \sum_{k-{\rm bins}} \frac{\left[ \Delta^2_{\rm 21,data}(k) - \Delta^2_{\rm 21,model}(k,\mathbf{p})\right]^2}{{\rm var}\left[ \Delta^2_{21}(k) \right]}\,,
\label{eq:likelihood}
\ee
where $\Delta^2_{\rm 21,model}$ is the model power spectrum defined in Eq.~\eqref{eq:vao_definition}, 
${\rm var}\left[ \Delta^2_{21}(k) \right]$ is the expected variance of the 21-cm power spectrum measurement for HERA 
which we generate using the \texttt{21cmSense} package 
for three different foreground scenarios as discussed in Section~\ref{sec:sensitivity}. 
The sum here is over the $k$-bins, $\mathbf{p}$ is the parameter vector which
we shall specify later. 
The small $k$-bins are mostly dominated by cosmic variance and foregrounds,
and the large $k$-bins are dominated by the telescope noise (see Figure~\ref{fig:pk_vao}).
The large $k$-modes also do not show VAOs. Based on these, we restrict the $k$-range
to $\{0.03,0.5\}\mpci$ for our analysis. We sample the likelihood space with the
Python package \texttt{emcee} \footnote{\href{https://github.com/dfm/emcee}
{github.com/dfm/emcee}} \cite{Foreman-Mackey:2012any}.


As discussed in Ref.~\cite{Munoz:2019fkt}, the current generation 21-cm observations 
will mostly measure the modes with $\kpa \gg \kpe$ due to the shape of the foreground wedge. 
As a consequence, 21-cm observations will not measure the AP parameter $\ape$ very precisely. 
Following Ref.~\cite{Munoz:2019fkt}, we keep  $\ape$ fixed during in our fitting and vary
$\apa$ in our MCMC analysis. However, we have checked that the inclusion of $\ape$
in the MCMC analysis does not affect our final results and we discuss this point in Section~\ref{sec:summary}.

We define our data vector $\mathbf{p}=\{ \apa, A_{\rm vel}, c_i\}$, where $c_i$ are the coefficients of
the model for the smooth part of the 21-cm power spectrum $\Delta_{21,\rm nw}^2(k,z)$ 
(Eq.~\eqref{eq:smooth_pw}), and $A_{\rm vel}$ is the
amplitude of the VAOs (Eq.~\eqref{eq:pk-VAO}). We impose the following priors on the parameters:
$ 0.8\leq \apa \leq 1.2$, $ 0 \leq A_{\rm vel} \leq 10^3\,{\rm mK}^2$, $-20 \leq c_i \leq 20$. These priors are
broad enough to fit the 21-cm power spectrum for all of the different simulations considered here. We find that
$n=2$ in Eq.~\eqref{eq:smooth_pw} is sufficient to model the smooth part of the 21-cm power spectrum 
for our $k$ range of fitting, and we
include coefficients $\{ c_0, c_1, c_2 \}$ in our analysis. Note that the parameters $(A_{\rm vel},c_0, c_1, c_2)$
depend mostly on the astrophysics. Therefore, marginalization over these parameters would mean marginalizing over the
astrophysical parameters.

\begin{table}[]
\resizebox{\columnwidth}{!}{ 
\begin{tabular}{|c|c|c|ccc|}
\hline
\multirow{2}{*}{Model} & \multirow{2}{*}{Feedback} & \multirow{2}{*}{$z_{\rm half}$} & \multicolumn{3}{c|}{$\Delta(Hr_d)/(Hr_d)\%$}                             \\ \cline{4-6} 
                   &                    &                    & \multicolumn{1}{c|}{Optimistic} & \multicolumn{1}{c|}{Moderate} & Pessimistic \\ \hline
\multirow{3}{*}{No Ly-$\alpha$ Heating} & No                  &           14.8         & \multicolumn{1}{c|}{0.3}  & \multicolumn{1}{c|}{0.62}  &  1.82 \\ \cline{2-6} 
                   & Low                  &       12.7             & \multicolumn{1}{c|}{0.51}  & \multicolumn{1}{c|}{1.14}  & 3.51  \\ \cline{2-6} 
                   & Regular                  &       12.2             & \multicolumn{1}{c|}{0.73}  & \multicolumn{1}{c|}{1.51}  & 4.62   \\ \Xhline{4\arrayrulewidth}
\multirow{3}{*}{With Ly-$\alpha$ Heating} & No                  &       15.1             & \multicolumn{1}{c|}{0.44}  & \multicolumn{1}{c|}{0.86}  &  2.64 \\ \cline{2-6} 
                   & Low                  &        13.3            & \multicolumn{1}{c|}{0.63}  & \multicolumn{1}{c|}{1.31}  &  3.94 \\ \cline{2-6} 
                   & Regular                  &         12.7           & \multicolumn{1}{c|}{0.9}  & \multicolumn{1}{c|}{1.84}  & 5.53   \\ \hline
\end{tabular}
}
\caption{This shows the projected $1-\sigma$ relative errors on $H(z) r_d$ for our fiducial parameter
set (Table~\ref{tab:parameters-fid}), under the different feedback, heating and foreground assumptions. 
We also mention the $z_{\rm half}$ values for each simulation. We have considered $540$ days of HERA observation
in all cases.} 
\label{tab:table1}
\end{table}

In Section~\ref{sec:epochs}, we discussed that the SNR for VAO measurement is expected to be 
maximum at $z_{\rm half}$ (defined for EoH). We restrict our analysis only to $z_{\rm half}$ which
is different for the different simulations considered here (see Table~\ref{tab:table1}). 
We run the MCMC analysis, based on the
likelihood in Eq.~\eqref{eq:likelihood} and the prior range discussed above, and present the
$1-\sigma$ marginalized relative error on $H r_d$ (obtained from the fitting of $\apa$) 
in Table~\ref{tab:table1}. We first consider the models without the \lya heating. 
For no LW feedback, $z_{\rm half}=14.8$ and we find that it is possible to measure $H r_d$
with $0.3\%$, $0.62\%$, $1.82\%$ accuracy for optimistic, moderate and pessimistic foreground 
contamination respectively. When we consider low and regular feedback, $z_{\rm half}$ is shifted
at lower value. Note that, the sensitivity of HERA is increased when we go to the low redshifts,
and we expect to have better measurement of $H r_d$ at low redshifts. However, as mentioned earlier,
the measurement of $H r_d$ (or VAOs) also depends on the amplitude of the VAOs. LW feedback of any form
disturbs star formation in small haloes, which not only delay the cosmic dawn, but also dampens the
VAO features. We actually see this in Table~\ref{tab:table1}. For low ($z_{\rm half}=12.7$) 
and regular ($z_{\rm half}=12.2$) feedback scenarios, 
it is possible to measure $H r_d$ with $(0.51,1.14,3.51)\%$ and $(0.73, 1.51, 4.62)\%$ relative accuracy for
the (optimistic, moderate, pessimistic) foregrounds. Now considering models with \lya heating, 
we overall see that $z_{\rm half}$ is higher and the constraints in each case are $\sim 1.5$ times worse
in comparison to models without \lya heating. We have already discussed in the previous section that 
\lya heating starts EoH earlier and make $z_{\rm half}$ higher where the sensitivity of HERA 
decreases. Not only that, \lya heating also decreases the amplitude of the 21-cm power spectrum, 
as well as of the VAOs. These two effects decreases the detectability of VAOs and thereby increase the 
relative error on the $H r_d$ measurement.

\begin{figure}
    \centering
    \includegraphics[width=1\columnwidth]{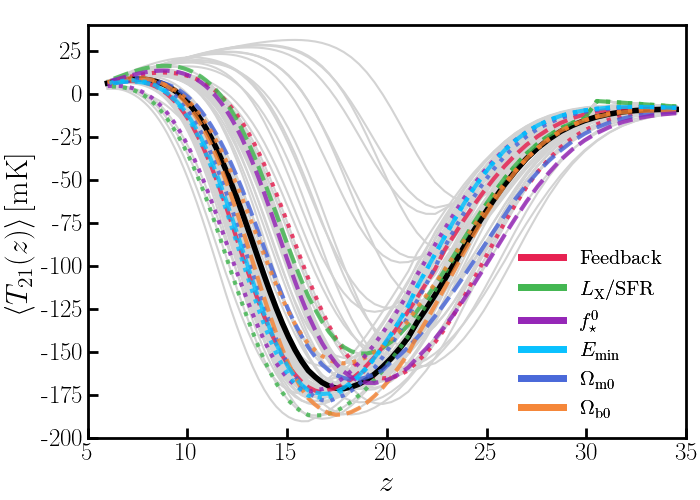}
    \caption{This shows the 21-cm global signals for different set of model parameters. We change the 
    various model parameters around the fiducial set (given in Table~\ref{tab:parameters-fid})
    and generate global signals for $\mathcal{O}(100)$ simulations. We show the fiducial signal, 
    which is for the fiducial parameters and low LW feedback, in thick solid-black line 
    and the rest of the global signals are plotted in thin solid-grey lines.
    To show the dependence of the signal on a few parameters, we highlight some of 
    the global signals in colours. The parameters and the associated colours are given in the legend. 
    The signals for parameter values above and below the fiducial are show respectively with dashed and 
    dotted lines. The parameters along with their values above and below the fiducial are as follows:
    Feedback=\{Regular, No\}, $\log_{10}(L_{\rm X}/{\rm SFR})=\{ 38.5, 39.5\}$, $f^0_{\star}=\{0.02,0.15\}$,
    $E_{\rm min}=\{ 0.5, 0.1\}$ keV,
    $\Omega_{\rm m0}=\{ 0.3311, 0.2911\}$, and $\Omega_{\rm b0}=\{ 0.0519, 0.0459\}$. Note that, for all the
    simulations we have considered \lya heating. 
    }
    \label{fig:many-globals}
\end{figure}

\begin{figure}
    \centering
    \includegraphics[width=1\columnwidth]{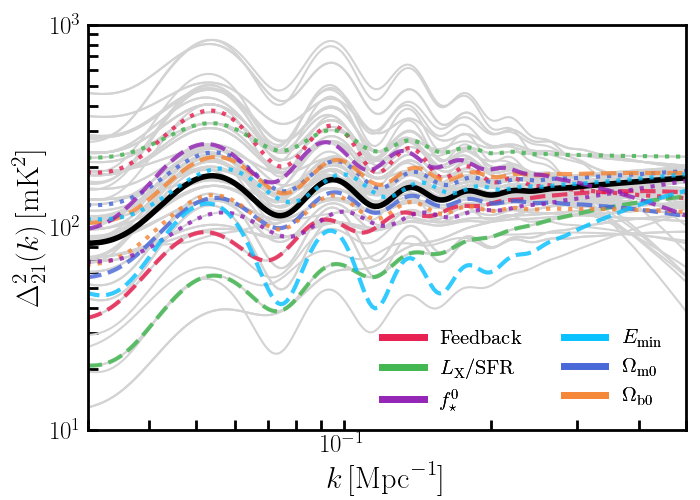}
    \caption{This shows the 21-cm power spectra at $z=z_{\rm half}$ for the same simulations shown in 
    Figure~\ref{fig:many-globals}. Different colours and line types carry the same meaning as in 
    Figure~\ref{fig:many-globals}. Note that for all the simulations,
    we have plotted the best fit 21-cm power spectra 
    (given by the model in Eq.~\eqref{eq:vao_definition}) 
    under the optimistic foreground assumption.}
    \label{fig:many-pk21s}
\end{figure}

\begin{figure*}
    \centering
    \includegraphics[width=1\textwidth]{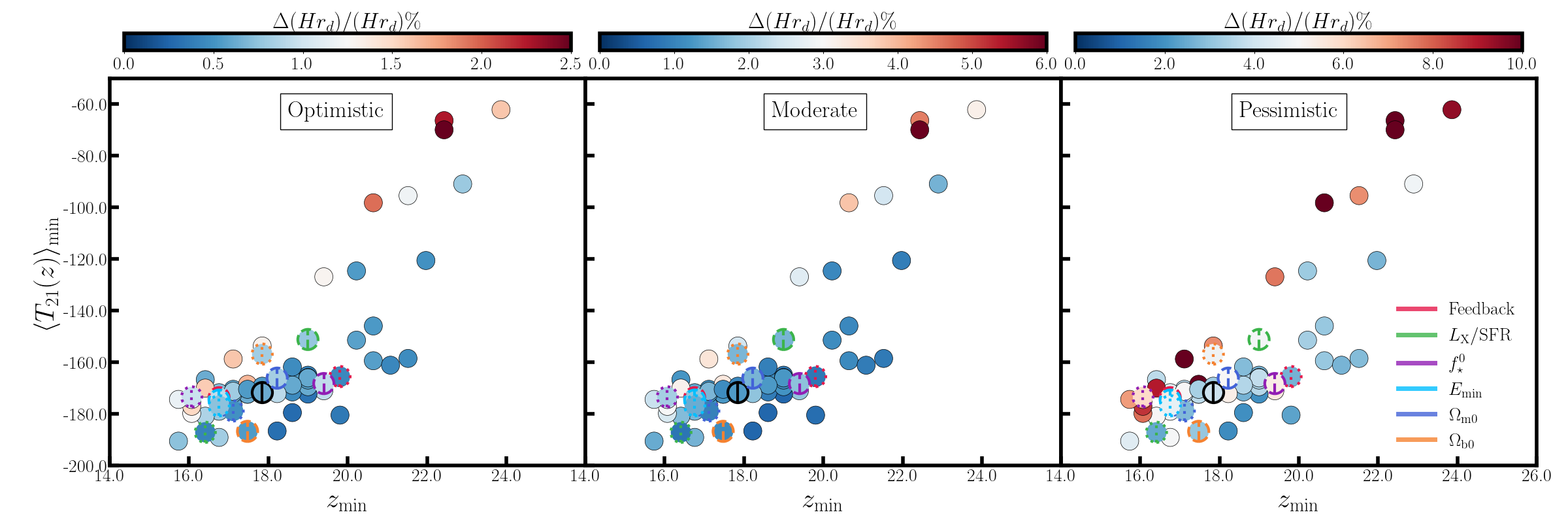}
    \caption{This shows the projected $1-\sigma$ relative errors on $H(z) r_d$ (obtained from the fitting of
    $\apa$ as discussed in Section~\ref{sec:measurement}) for various simulations 
    mentioned in Figures~\ref{fig:many-globals} and \ref{fig:many-pk21s}. 
    Each simulation here is denoted by its corresponding 
    ($z_{\rm min},\langle T_{21} \rangle_{\rm min}$) values, which we
    show as circles, while the colour on the circles show
    the $1-\sigma$ relative errors on $H r_d$. We have shown results for three foreground removal scenarios:
    optimistic, moderate and pessimistic. The circles corresponding to the highlighted parameters in Figures~\ref{fig:many-globals} and \ref{fig:many-pk21s} are shown here with coloured edges and 
    central bars. The different colours of the edges and the bars represent the parameters 
    (given in the legend), while the dashed and dotted line styles respectively represent values above and below 
    the fiducial.
    }
    \label{fig:ap_constraints}
\end{figure*}

\begin{figure}
    \centering
    \includegraphics[width=1\columnwidth]{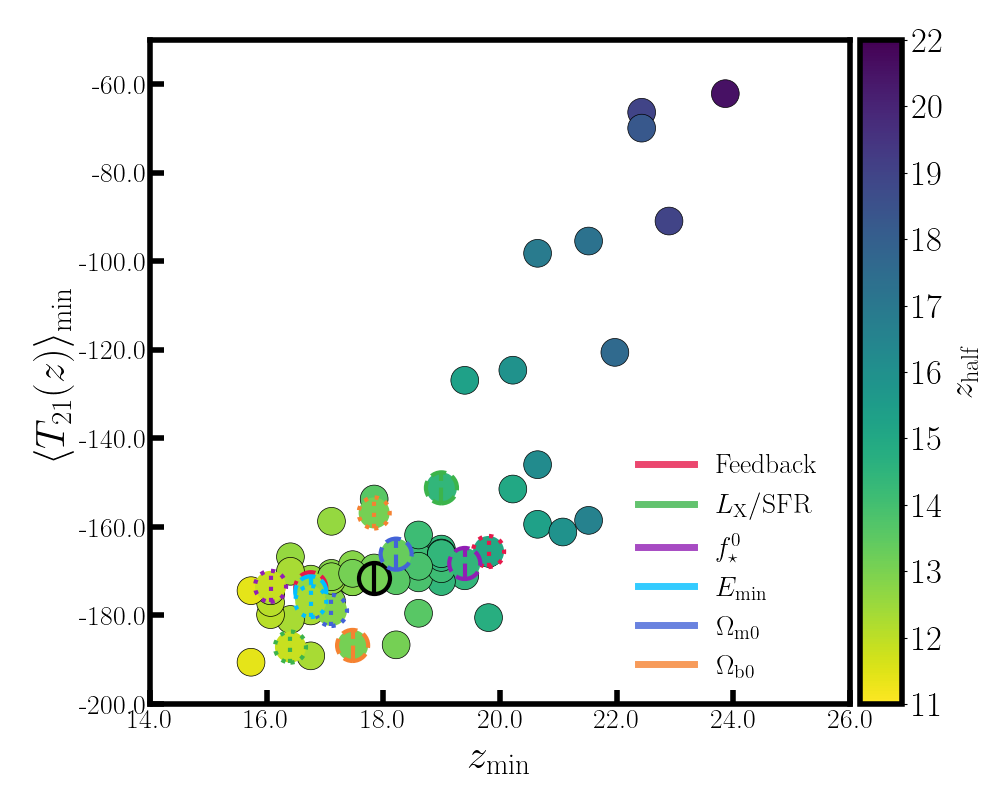}
    \caption{For the same circles shown in Figure~.\ref{fig:ap_constraints}, different colours here
    show the $z_{\rm half}$ values for the simulations denoted by their 
    ($z_{\rm min}, \langle T_{21} \rangle_{\rm min}$) values.}
    \label{fig:zhalf}
\end{figure}

Note that, Ref.~\cite{Munoz:2019fkt} has performed the analysis 
at two redshifts $z=16$ and $18$ for different simulations. 
Their analysis helped to compare the
measurements of $H r_d$ at each $z$ for different simulation scenarios. 
However, our aim here is different. 
The astrophysics is largely unknown at high redshifts and $z_{\rm half}$ depends 
very much on the model of astrophysics (as well as cosmology).
We want to quantify the best possible measurement of $H r_d$
(which occurs close to $z_{\rm half}$) for all the different astrophysical 
and cosmological scenarios.
To do so, we run $\mathcal{O}(100)$ different simulations by changing the model parameters
around the fiducial set (which ultimately change $z_{\rm half}$), 
and perform the fitting process discussed above to determine the 
$1-\sigma$ relative error on $H r_d$ at $z_{\rm half}$. We have changed the following parameters and
we mention the range, over which the parameters are changed, in the curly brackets. 
The parameters are: $\zeta=\{18,22\}$, $\lambda_{\rm MFP}=\{10,20\}\,{\rm Mpc}$, 
$V^{(0)}_{\rm cool}=\{3.5,4.5\}\,{\rm km/s}$,  
$\log_{10} \left( L_{\rm X}/{\rm SFR} \right)=\{38,41\}$, $\alpha_X=\{1.0,1.5\}$,
$f^0_{\star}=\{0.02, 0.15\}$, $E_{\rm min}=\{0.1,0.5\}$ keV, $\Omega_{\rm m0}=\{0.2911, 0.3311\}$,
and $\Omega_{\rm b0}=\{0.0459, 0.0519\}$. Note that for each set of parameters, we have considered three
LW feedback strengths. We also include \lya heating in all the simulations.  
For comparison, we plot the 21-cm global signals and power spectra for all the simulations 
in Figures~\ref{fig:many-globals} and \ref{fig:many-pk21s} respectively.
Instead of denoting each simulation with their corresponding parameter sets, we
denote them with their corresponding $\langle T_{21} \rangle_{\rm min}$ and $z_{\rm min}$ values
(which are different for the different set of parameters) and
present the fitting results in Fig.~\ref{fig:ap_constraints}. 
We also plot the corresponding $z_{\rm half}$ values for each simulation in Fig.~\ref{fig:zhalf}. 
We discuss our fitting results in the following paragraph.

From Figure~\ref{fig:ap_constraints}, we see that the 
$z_{\rm min}$ and $\langle T_{21} \rangle_{\rm min}$ values are positively correlated
for the range of parameters that we have chosen.
Considering Figure~\ref{fig:many-globals}, we find that models e.g. with 
low LW feedback, high X-ray heating, large $f^0_{\star}$, large $\Omega_{\rm m0}$,
start the EoH early and we see higher $\langle T_{21} \rangle_{\rm min}$ and $z_{\rm min}$ values for these models.
The corresponding $z_{\rm half}$ values also increase for these models (with respect to the 
fiducial model) as can be seen from Figure~\ref{fig:zhalf}. Considering the overall results 
from Figure~\ref{fig:ap_constraints}, we find that for $z_{\rm min}<22$ ($z_{\rm half}<16$) and 
$\langle T_{21} \rangle_{\rm min}<-100$ mK, measurement of $Hr_d$ is possible with 
$\lesssim 1.5\%$, $\lesssim 3\%$ and $\lesssim 6\%$ accuracy 
for optimistic, moderate and pessimistic foreground scenarios. At $z_{\rm min}>22$, we see that the
errors somewhat increase and we find that most models have high X-ray heating in this range. 
We also see that for a fixed LW feedback strength, the errors mostly 
decrease as we go to smaller $z_{\rm min}$ (or $z_{\rm half}$). As discussed previously, sensitivity of HERA is 
increased as we go to lower redshifts and this helps to reduce the errors in the measurement of $Hr_d$.
The error also depends on other factors, which we shall discuss below.

We now discuss the effects of the parameters, highlighted with colours in 
Figure~\ref{fig:ap_constraints}, 
in the measurement of $Hr_d$. Note that, the effects will be discussed with respect to the fiducial
parameter set. Also, we shall take help from Figures~\ref{fig:many-globals} and \ref{fig:many-pk21s} 
at each step in order to explain and interpret  the results plotted in Figure~\ref{fig:ap_constraints}.
We have already discussed the effects of the LW feedback, increasing which 
delays the EoH (See Figure~\ref{fig:many-globals}), 
the VAOs tend to get erased (See Figure~\ref{fig:many-pk21s}) and ultimately the errors 
in $H r_d$ are increased (Table~\ref{tab:table1}). 
Considering the X-ray heating parameter $\log_{10} \left( L_{\rm X}/{\rm SFR} \right)$, 
we find that EoH occurs early (Figure~\ref{fig:many-globals}) and $z_{\rm min}$ (or $z_{\rm half}$)
is higher when we increase $\log_{10} \left( L_{\rm X}/{\rm SFR} \right)$ with respect to the fiducial value
and vice versa. Higher X-ray heating also damps the VAO amplitude as can be seen from 
Figure~\ref{fig:many-pk21s}. If we combine both effects, we see that the measurement errors increase and
decrease when $\log_{10} \left( L_{\rm X}/{\rm SFR} \right)$ is respectively increased and decreased.
Next we discuss the effects of $f^0_{\star}$ which controls the fraction of baryons into stars. Increasing
and decreasing $f^0_{\star}$ from the fiducial value make  $z_{\rm min}$ (or $z_{\rm half}$)
respectively higher and lower (Figure~\ref{fig:many-globals}). Not only that, the VAO amplitude is also
higher for the models with high $f^0_{\star}$ and vice versa (Figure~\ref{fig:many-pk21s}). 
These two competing effects decide the measurement errors of $Hr_d$. For the values 
$f^0_{\star}=$ 0.02 and 0.15, which we have highlighted in Figure~\ref{fig:ap_constraints},
we find that the error is slightly lower for the model with $f^0_{\star}=0.15$, although  
$z_{\rm half}$ is considerably higher for this model. This again shows the importance of the VAO amplitude
to decide the signal-to-noise in $Hr_d$ measurement. 
Next we consider $E_{\rm min}$. We see that higher $E_{\rm min}$ increases $z_{\rm min}$ (or $z_{\rm half}$)
(Figure~\ref{fig:many-globals}) and damps the VAOs at large $k$ (Figure~\ref{fig:many-pk21s}). Opposite
happens for lower $E_{\rm min}$. As a result, measurement errors of $Hr_d$ increases for high 
$E_{\rm min}$ models and vice versa (Figure~\ref{fig:ap_constraints}).
Considering $\Omega_{\rm m0}$, we find that 
CD and subsequent epochs start early for models with high $\Omega_{\rm m0}$, which also increases 
$z_{\rm min}$ (or $z_{\rm half}$) (Figure~\ref{fig:many-globals}). 
However, high $\Omega_{\rm m0}$ dampens the VAO features, and VAOs are 
more pronounced for models with low $\Omega_{\rm m0}$ (Figure~\ref{fig:many-pk21s}).
Combination of these two effects make $Hr_d$ measurement error high for models with high $\Omega_{\rm m0}$,
and the opposite is also true. The effect of $\Omega_{\rm b0}$ is not straightforward to guess. 
Naively once expects that
$z_{\rm min}$ will be higher for models with high $\Omega_{\rm b0}$, as the CD and EoH are expected to start early 
for these models. However, we find that, $\Tk$ is lower for the models with high $\Omega_{\rm b0}$. Also 
the coupling between $\Ts$ and $\Tk$ is higher for these models. As a result, we find that both 
$z_{\rm min}$ (also $z_{\rm half}$) and $\langle T_{21} \rangle_{\rm min}$ are slightly lower for the model with high 
$\Omega_{\rm b0}$ (Figure~\ref{fig:many-globals}). Also, high $\Omega_{\rm b0}$ boosts VAO amplitude
(Figure~\ref{fig:many-pk21s}). All these effects together help in reducing the error in $Hr_d$
for models with high $\Omega_{\rm b0}$. Overall, we can conclude that, the measurement errors of 
$H r_d$ are smaller for models that produce lower
$z_{\rm min}$ (or $z_{\rm half}$) and $\langle T_{21} \rangle_{\rm min}$, without damping the VAOs considerably.

\subsection{Errors due to the uncertainties in cosmological parameters}
\label{sec:degeneracy_cosmological}

\begin{table}[h]
\resizebox{\columnwidth}{!}{
\begin{tabular}{|c|c|c|c|c|}
\hline
               Model   & Feedback & $z_{\rm half}$ & \makecell{$\left(\frac{\Delta \apa}{\apa}\right)\%$ \\ TT+lowE} & \makecell{$\left(\frac{\Delta \apa}{\apa}\right)\%$ \\ TT,TE,EE\\+lowE+lensing\\+BAO} \\ \hline
\multirow{3}{*}{With Ly-$\alpha$ Heating} & No & 15.1 &	0.16 &	0.081  \\ \cline{2-5} 
                  & Low & 13.3 &	0.17 &	0.086 \\ \cline{2-5} 
                  & Regular &  12.7 &	0.17 &	0.084  \\ \hline
\end{tabular}
}
\caption{This shows the relative errors introduced in the measurement of $\apa$ by the 
uncertainties in the Planck-2018 cosmological parameters \cite{Planck:2018vyg}
for our fiducial parameter set (Table~\ref{tab:parameters-fid}), under the different feedback scenarios.
We have used two Planck-2018 data sets, namely `TT+lowE' and 
`TT,TE,EE+lowE+lensing+BAO', to calculate the relative errors as discussed in 
Section~\ref{sec:degeneracy_cosmological}.
We also mention the $z_{\rm half}$ values for each simulation.}
\label{tab:table2}
\end{table}

The cosmological parameters are very precisely constrained by the Planck measurements \cite{Planck:2018vyg}. 
For this reason, we often do not consider the uncertainties in the cosmological parameters, 
when we study the 21-cm signal. We assume that
most of the uncertainties are introduced by the astrophysical parameters. 
However, we have already seen that for optimistic foreground contamination, measurement
of $H r_d$ is possible with sub-percent precision. In this scenario, we must check how much does the 
uncertainty in cosmological parameters influence the  measurement of $H r_d$ (or $\apa$). 
However, it is not straight forward to infer this. Here, we follow the technique introduced in 
Ref.~\cite{Liu:2015txa} to determine this uncertainty. $\apa(\mathbf{p})$ is a function of 
cosmological parameters and we denote the parameter vector by $\mathbf{p}$, where 
$\bar{\mathbf{p}}$ represents its fiducial value. 
The fiducial values are given in Table~\ref{tab:parameters-fid}.
We now consider a small variation around the fiducial parameter
values $\bar{p}_m$ and assume the following linearised relation
\be
    \left( \frac{\Delta \apa}{\apa} \right) (\bar{\mathbf{p}})= \sum_{m} \left( \frac{\partial \ln \apa}{\partial p_m} \right) \Delta p_m \,,
    \label{eq:degen}
\ee
where the index $m$ denotes the different cosmological parameters, and the coefficients 
$ \frac{\partial \ln \apa}{\partial p_m} (\bar{\mathbf{p}})$ 
capture the variation in $\left( \Delta \apa/\apa \right)$ due to the variation $\Delta p_m$
in any cosmological parameters. We calculate these coefficients from simulations. We consider the 
21-cm power spectra at $z=z_{\rm half}$ for each set of simulations discussed in Table~\ref{tab:table1}.
We then perform the fitting process discussed in section~\ref{sec:measurement} to determine the 
parameters $\{ \apa, A_{\rm vel}, c_0, c_1, c_2\}$ for the same range $k \in\{0.03,0.5\}\mpci$
as used in Section~\ref{sec:measurement}. 
We have considered only the Poisson errors at each 
$k$-bin (shown in Fig.~\ref{fig:pk_vao}) during the fitting, and ensure that we get best fit $\apa \approx 1$
which is what we expect to get for a correct set of cosmology. 
In this fitting process, we also checked that the errors on the best fit parameters are small enough that they 
do not have much influence on the results that we discuss next. After fitting the parameters for the fiducial
simulation set, we consider one cosmological parameter and take two values of it at $\pm \delta p_m$ away from the
fiducial value $\bar{p}_m$ and run the simulations. We choose the $\delta p_m$ values such that the parameters 
$\{A_{\rm vel}, c_0, c1, c2\}$ stay close to the values for the fiducial simulation. 
Next, we seek the best fit $\apa$
value for simulations with $(\bar{p}_m \pm \delta p_m)$ at the same $z_{\rm half}$
as the fiducial simulations, while keeping 
$\{A_{\rm vel}, c_0, c1, c2\}$ values fixed. We also use the fiducial parameter set to determine
$H^{\rm fid}(z)$ and $r^{\rm fid}_d$ values. This process ensures that the change $\delta p_m$ will 
only change the $\apa$ values. After getting the two $\apa$ values at $(\bar{p}_m \pm \delta p_m)$, we calculate the
coefficient $ \frac{\partial \ln \apa}{\partial p_m} (\mathbf{p})$.
We apply the same process for all the cosmological parameters, and find that only 
$\Omega_{\rm m0}$ and $\Omega_{\rm b0}$ change the $\apa$ values significantly, which is expected.
Like in section~\ref{sec:measurement}, we also do not consider the variation of the parameter $\ape$,
while performing the fitting process.
Note that, the uncertainties $\Delta p_m$ in Eq.~\eqref{eq:degen}
are from Planck measurement. In order to determine the
relative uncertainty $(\Delta \apa / \apa)$ due to $\Delta p_m$, we do the following.
We draw $\mathcal{O}(10^5)$ samples for the set ($\Omega_{\rm m0}$, $\Omega_{\rm b0}$) from different
Planck-2018~\cite{Planck:2018vyg}  likelihoods,
determine $(\Delta \apa / \apa)$ for each set using Eq.~\eqref{eq:degen}
and get a distribution of $(\Delta \apa / \apa)$ values. The 
$1-\sigma$ uncertainty in $(\Delta \apa / \apa)$ is then determined from the standard deviation of this 
distribution. Note that, we have drawn the random samples until we reach a stable Gaussian distribution
for $(\Delta \apa / \apa)$ with a fully converged standard deviation. We find that, for all the different simulations
considered in Table~\ref{tab:table2}, the uncertainties in $\Omega_{\rm m0}$ and  $\Omega_{\rm b0}$ 
from Planck-2018 introduce $\sim 0.08-0.2\%$ relative error in $H r_d$ measurement. 
This is significant if we consider the relative error values
for optimistic and even for moderate foregrounds in Table~\ref{tab:table1}. 
Note that, the relative errors in Table~\ref{tab:table2} provide an error floor, which can be
minimised only with more precise measurements of the cosmological parameters, which is possible
with the next generation of experiments, like {\it CMB stage-4} \cite{CMB-S4:2022ght}, {\it EUCLID} 
galaxy survey \cite{EuclidTheoryWorkingGroup:2012gxx}, etc.

\section{Summary and Conclusions}
\label{sec:summary}

The fluctuations in the DM-baryon relative velocity field are believed to be imprinted as 
well predicted acoustic oscillations called VAOs in the 21-cm power spectrum at cosmic dawn and 
also possibly at early reionization. These VAOs can be treated as a standard
ruler to measure the expansion rate $H(z)$ at high redshifts. In this paper, we have simulated the
21-cm signal that includes the DM-baryon relative velocity fluctuation effects. 
We also model the VAOs in the 21-cm power spectrum using analytical prescriptions. 
We perform Alcock-Paczy\'{n}sky (AP) tests on the simulated data, 
and quantify the relative error with which the combination $H(z)r_d$ (or $\apa$ parameter)
can be measured under the different heating, LW feedback and foreground scenarios. 
We consider the 21-cm observations with the HERA interferometer and assume three foreground
removal scenarios: optimistic, moderate and pessimistic. 
We contained our analysis only to EoH, which occurs at $z<z_{\rm min}$ (where $z_{\rm min}$ 
refers to the redshift where the global signal $\langle T_{21} \rangle = \langle T_{21} \rangle_{\rm min}$ has its minimum
value), and measure $H(z)r_d$ at $z=z_{\rm half}$ where 
$\langle T_{21} \rangle = (1/2) \langle T_{21} \rangle_{\rm min}$ and the signal-to-noise for VAO 
measurement is highest close to this redshift. The $z_{\rm half}$ values change with changes in 
heating and LW feedback scenarios, and the change $\Delta z_{\rm half}$ can be as large as $\sim 3$
between the different scenarios. We quote our main results (for the fiducial set of parameters given in
Table~\ref{tab:parameters-fid}) in Table~\ref{tab:table1}. 
We find that, if no LW feedback is present, $H(z)r_d$ can be measured with $0.3, 0.62$ and $1.82\%$
relative accuracy for optimistic, moderate and pessimistic foreground removal scenarios respectively. 
LW feedback tend to erase the VAO peaks and, as a result, the relative accuracy in $H(z)r_d$
measurement decreases. When we consider \lya heating in our simulations, which reduces the 
21-cm power spectrum amplitude and thereby VAO peaks, we find that the errors are $\sim 1.5$ times worst 
when compared against the simulations with no \lya heating. This is a significant effect, depending
on the strength of the X-ray heating. 

To quantify the best possible measurement of $H r_d$ for
all the different astrophysical and cosmological scenarios, we run $\mathcal{O}(100)$ different simulations by
changing the model parameters around the fiducial set
which change the $z_{\rm min}$ and  $\langle T_{21} \rangle_{\rm min}$ values. We find that 
for $14<z_{\rm min}<22$ (or $11<z_{\rm half}<16$), it is possible to measure $H(z)r_d$ with $\lesssim 1.5\%$,
$\lesssim 3\%$ and $\lesssim 6\%$ accuracy for optimistic, moderate and pessimistic foregrounds respectively.
These results are in agreement with Ref.~\cite{Munoz:2019fkt}.
At $z_{\rm min}>22$, errors increase substantially. We checked the errors introduced by the 
uncertainty in cosmological parameters from Planck satellite measurements, and found that the uncertainties in 
$\Omega_{\rm m0}$ and $\Omega_{\rm b0}$ introduce $\sim 0.08-0.2\%$ relative error in $H(z)r_d$
measurement. Considering the errors for optimistic and even for moderate foregrounds, this is significant.

Considering the Local measurements, like SHOES \cite{DiValentino:2020zio}, and the galaxy surveys, like 
Baryon Oscillation Spectroscopic Survey (BOSS) \cite{Ivanov:2019pdj}, 
we have measurement of $H(z)$ up to $z \lesssim 2$. 
Future galaxy surveys, like \textit{Euclid} \citep{Euclid:2021qvm,Mazumdar:2022ynd} 
and \textit{SPHEREX} \cite{Dore:2014cca}, are expected to measure $H(z)$
up to a maximum redshift of $z \sim 5$. Line intensity mapping (LIM) using 
the post reionization 21-cm signal, we can reach 
up to $z\sim6$ \cite{Sarkar:2016lvb,Sarkar:2018gcb,Pourtsidou:2016dzn,Obuljen:2017jiy,Olivari:2017bfv,
Chakraborty:2020zmx}. LIM technique using other lines, like:
carbon monoxide (CO) rotational lines 
\cite{Lidz:2011dx,Pullen:2012su,Breysse:2014uia,Padmanabhan:2017ate,Libanore:2022ntl},
[CII] \cite{Silva:2014ira,Pullen:2017ogs,Padmanabhan:2018yul}, 
$H\alpha$ and $H\beta$ \cite{Silva:2018}, oxygen lines \cite{Gong:2018}, \lya \cite{Pullen:2013dir}, etc.,
can measure $H(z)$ up to $z\lesssim9$ \cite{Bernal:2019jdo,Bernal:2019gfq}. 
Our current work predicts that, using VAOs in the 21-cm power spectrum, it is possible to
measure $H(z)$ with reasonable accuracy in the range $10 < z < 20$. The redshift range can be extended
further if we include the VAOs from LCE ($z>20$). However, we have analysed the VAO signal from LCE 
and found that the current generation of 21-cm experiments, like HERA, are not very sensitive to
the measurement of $H(z)$ from this epoch.

Our analysis has a few limitations which we discuss below. 
As the current generation 21-cm observations 
will mostly measure the modes with $\kpa \gg \kpe$ due to the shape of the foreground wedge,
we vary $\apa$ in our analysis while we keep $\ape$ fixed throughout.
However, it is also possible to include the variation of $\ape$ in the analysis 
(as described in Ref.~\cite{Munoz:2019fkt}), and this will provide a constraint on 
the measurement of $D_A(z)$. We have performed a separate analysis with both $\apa$ and $\ape$
parameters for a few simulations mentioned in Table~\ref{tab:table1}
and we got similar results as found by Ref.~\cite{Munoz:2019fkt}. The constraints on $D_A(z)$ are
not very good except for the optimistic foreground case. 
We note that the relative errors for $H(z) r_d$ in Table~\ref{tab:table1} are largely 
unaffected by the inclusion of $\ape$.
The VAO amplitude is also affected by the
additional feedback effects like:  galactic winds emanating from star-forming regions  \cite{Springel:2002ux},
photo evaporation of small galaxies \cite{Barkana:1999apa}, ultraviolet radiative feedback 
\cite{Mesinger:2008ze}, etc.,
which hinder the star formation in different haloes, and we 
do not include these in our analysis. Presence of exotic dark matters, 
like \textit{warm} \cite{Sitwell:2013fpa,Boyarsky:2019fgp,Munoz:2019hjh} and \textit{fuzzy} \cite{Jones:2021mrs,Hotinli:2021vxg,Sarkar:2022dvl}, also 
affect the formation of mini haloes and thereby affect the VAO signature in the 21-cm power spectrum. 
Spectrum of the early X-ray sources also determine the detectability of VAOs \cite{Pacucci:2014wwa,Fialkov:2014kta}. 
Recently, Ref.~\cite{Munoz:2021psm} has pointed out that the parametrization for 
star formation rate density used in \texttt{21cmvFAST} code 
does not reproduce the stellar-to-halo mass relation observed in UV Luminosity Functions from HST.
Further, \texttt{21cmvFAST} assumes isotropic LW feedback which may overpredict VAOs 
(see Ref.~\cite{Munoz:2021psm}). Our analysis is done for coeval boxes. VAO amplitude changes 
if the power spectra are calculated for light-cone boxes \cite{Munoz:2021psm}, which are more realistic. 
We leave all these for future study.

We finally conclude that the detection of the VAO features in the 21-cm signal using HERA radio telescope 
enables us to measure the expansion rate $H(z)$ of the universe during the cosmic dawn and early 
reionization era ($11 < z <20$) with reasonable accuracy. The precision with which $H(z)$ can be 
measured is comparable to those of the low redshift measurements.
This can be helpful in testing various cosmological scenarios, and also can be helpful in alleviating the 
ongoing ``Hubble Tension".

\appendix

\begin{acknowledgements}
We thank Julian Mu\~{n}oz for useful discussions and comments on the manuscript. 
EDK acknowledges support from an Azrieli faculty fellowship.
\end{acknowledgements}

\end{document}